# Optical Properties of Amorphous Silicon Quantum Dots (a-Si QDs) with various dot size using Extended Hückel Theory

K.M. Liu[1] and Setianto[1]

**Abstract.** A high quality amorphous silicon (a-Si) nanostructures has grown experimentally to study the origin of light emission and the quantum confinement effect in a-Si. The quantum confinement effect increases the band gap of material as the size of quantum structure decreases, which results in a blue shift in optical luminescence and energy absorption. Here we demonstrate this effect using extended Hückel method to calculate fundamental band gap and optical absorption energy of a-Si samples with various dot sizes. As result, when the dot size was decreased from 2.2 to 1.0 nm, the absorption spectra peak shifted toward higher energy from 2.278 eV to 3.856 eV.

## 1 Introduction

Amorphous silicon (a-Si) has two important advantages compared with bulk crystalline Si. The luminescence efficiency in bulk a-Si is higher than that in crystalline Si due to structural disorder and the band gap energy of bulk a-Si (1.6 eV) is larger than that of bulk crystalline Si (1.1 eV) and a-Si is a good candidate for short-wavelength luminescence. It is, therefore, expected that these intrinsic advantages of a-Si and quantum confinement effect of a-Si quantum dot (a-Si QD) lead to the extremely useful optical properties in the short wavelength region compared with porous Si or nano – crystalline Si [R. A. Street (1981): G. Allan, C. Delerue, and M. Lannoo (1997): M. J. Estes and G. Moddel(1996)]. In addition, the study of the quantum confinement effect in a-Si as a function of dimension can yield important information about the physical processes related to the carrier mobility in a-Si. In contrast to crystalline Si, little is known about the limiting processes of transport in a-Si. Therefore, it is very important to grow high quality a-Si nanostructures to experimentally study the origin of light emission and the quantum confinement effect in a-Si.

## 2 Methodology

In molecular orbital (MO) theory, the electronic structure of a molecule is described by molecular orbitals $\psi_i$ (i=1,2,…m) that are expressed as a linear combination of atomic orbitals $\chi_\mu$ (μ=1,2,…,m),

$$\Psi_i = \sum_\mu C_{\mu i} \chi_\mu \qquad (1)$$

---

[1] Department of Physics, Faculty of Mathematics and Natural Sciences, Universitas Padjadjaran Jalan Raya Bandung-Sumedang KM 21, Jatinangor 45363, West Java, Indonesia
email: setianto@phys.unpad.ac.id

where $C_{\mu i}$ is the coefficient of the atomic orbital $\chi_\mu$. The energies $e_i$ of the MO's $\Psi_i$ are eigen values associated with the effective Hamiltonian $H^{eff}$

$$H^{eff}\Psi_i = e_i \Psi_i \quad (2)$$

So that $e_i$ is written as

$$e_i = \frac{\langle \Psi_i | H^{eff} | \Psi_i \rangle}{\langle \Psi_i | \Psi_i \rangle} \quad (3)$$

The energy $e_i$ is optimum with respect to changes in the coefficients $C_{Ki}$ so that

$$\frac{\partial e_i}{\partial C_{Ki}} = 0 \quad (K = 1,2,\ldots,m) \quad (4)$$

which leads to the m x m secular determinant

$$\begin{vmatrix} H_{11}-e_i S_{11} & H_{12}-e_i S_{12} & \cdots & H_{1m}-e_i S_{1m} \\ H_{21}-e_i S_{21} & H_{22}-e_i S_{22} & \cdots & H_{2m}-e_i S_{2m} \\ \vdots & \vdots & \vdots & \vdots \\ H_{m1}-e_i S_{m1} & H_{m2}-e_i S_{m2} & \cdots & H_{mm}-e_i S_{mm} \end{vmatrix} = 0 \quad (5)$$

where the matrix elements $H_{\mu v}$ and $S_{\mu v}$ are expressed as

$$H_{\mu v} = \langle \chi_\mu | H^{eff} | \chi_v \rangle \quad (6)$$

$$S_{\mu v} = \langle \chi_\mu | \chi_v \rangle \quad (7)$$

The matrix elements $H_{\mu v}$, $S_{\mu v}$ and $C_{\mu i}$ ($\mu,v,i = 1, 2, \ldots,m$) form m x m matrices **H**, **S** and **C**, respectively. These matrices are related by the pseudo-eigenvalue equation,

$$\mathbf{HC} = \mathbf{SCe} \quad (8)$$

where **e** is an m x m diagonal matrix with its diagonal elements given by $e_i$ (i= 1,2, …, m). Thus is the matrices **H** and **S** are determined, and then the coefficient matrix **C** and orbital energies **e** are obtained by solving the pseudo-eigenvalue problem, Eq. 8.

The crucial difference between first principles and semi-empirical methods of electronic structure calculations lies in the way of constructing the **H** matrix using the Hamiltonian appropriate for the system and refine it in a self-consistent field manner. In semi-empirical methods, the construction of H is carried out empirically by introducing a number of simplifying approximations.

In the Extended Hückel Theory (EHT) method, only valence electrons are considered, and valence atomic orbitals are approximated by Slater type orbitals (STO's) $\chi_\mu$ [Hoffmann, R. J. (1963)]. Single-zeta STO's are defined by

$$\chi_\mu(r,\theta,\phi) \propto r^{n-1} \exp(-\zeta r) Y(\theta,\phi) \quad (9)$$

where n is the principal quantum number, $\zeta$ is the exponent, and $Y(\theta,\phi)$ is the spherical harmonics. In EHT method the exact mathematical forms of the effective Hamiltonian $H^{eff}$ is empirically constructed. The values of valence orbitals can be approximated by their orbital energies from suitable atomic electronic structure calculations [Clementi, E.; Roetti, C.(1974)].

The energy gap between valence and conduction band is of fundamental importance for the properties of a solid. Most of a material's behavior, such as intrinsic conductivity, optical transitions, or electronic transitions, depends on it [McLean, A. D.; McLean, R. S(1981)]. Any change of the gap may significantly alter the material's physics and chemistry. This occurs when the size of a solid is reduced to the nanometer length scale. Therefore, the science and the technology of nanomaterial need to take into account a band gap, which is different from that of the bulk.

The band gap for semiconductor quantum dots is usually quite well described by an extended effective mass approximation

(EMA). This describes a band gap, which gradually increases for smaller sized particles. For very small clusters of semiconductors, the EMA does apply. The basic approach to cluster properties is to start from the atom, and calculate the gap with increasing cluster size. For clusters containing just a few atoms, surface passivation would change their intrinsic properties significantly. Therefore, the properties of a pristine cluster have to be studied with the assumption of a bare surface. Most often, the clusters do not have the bulk atomic structure. One finds for instance that not the diamond structure but rather the close-packed structure gives the global minimum of small silicon clusters. A covalent-metallic transition is predicted which leads to a band gap for silicon clusters much smaller than the 1.1 eV bulk gap.

## 3 Results and Discussion

A nanoparticle of a- Si QD's with a diamond crystal structure is shown in Fig 1. For silicon particles, a structural transformation is predicted for smaller sizes [Richardson, J. W.; Blackman, M. J.; Ranochak, J. E(1973)]. In these studies, Si-$n$ cluster growth follows diamond structure up to $n = 200$ (ca. 2 nm diameter). Obviously, simple models like the effective mass approximation are not applicable to pristine silicon clusters. Si-$n$ with just a few atoms ($n < 10$) was theoretically found to have compact cub octahedral or icosahedral structure. The average coordination number for silicon clusters with $n > 29$ tends to lie above the coordination number 4 of the bulk diamond structure. As bulk silicon is a semiconductor, a major change in the electronic properties can therefore be expected with size reduction. HOMO-LUMO energy gap several electron volts were predicted for small silicon clusters.

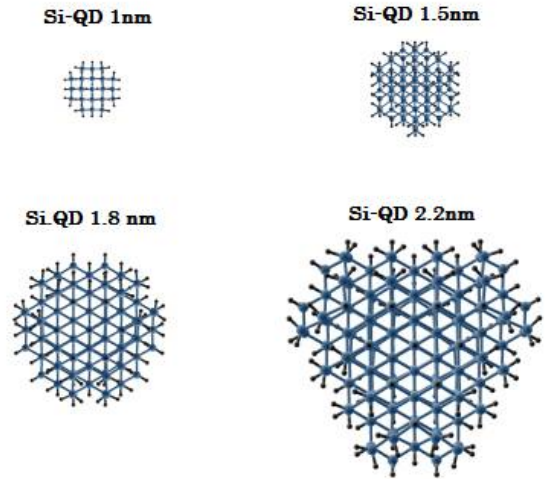

Fig. 1. Model of Si-QD with a diamond-type crystal structure.

For example, we obtained 3.86 eV for Si-QD 1nm, while gaps of 2.75 eV for Si-QD 1.5nm. A detail EHT calculation is tabulated in Table 1.

The quantum dot size effect predicts the formation of a band gap with decreasing particle size of Si-QD's is shown in Fig 2.

**TABLE 1.** Calculated optical band gap energies for silicon cluster dependent on their diameter

| Si-QD | HOMO-LUMO energy gap (eV) |
|---|---|
| Si-QD 1nm | 3.86 |
| Si-QD 1.5nm | 2.75 |
| Si-QD 1.8nm | 2.45 |
| Si-QD 2.2nm | 2.28 |

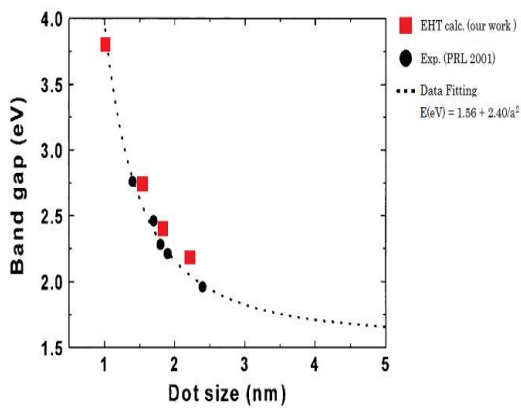

Fig. 2. a-Si QDs as a function of the dot size.

## 4 Conclusion

We have presented EHT calculations for investigating optical properties of Si-QDs, i.e. HOMO - LUMO energy gap with different dot size. There is direct relation between energy gap with dot size, Si-QDs in this case. Need to study further for different atom/ligand terminated on Si-QDs such as Oxygen, Nitride and Carbon atoms, etc.

**Acknowledgement** We would like to thank Professor R.E. Siregar for short course of the computational method in chemistry.